\documentclass[english,aps,prb,showpacs]{revtex4}
\usepackage[T1]{fontenc}
\usepackage[latin9]{inputenc}
\usepackage{mathrsfs}
\usepackage{amsmath}
\usepackage{amssymb}

\makeatletter
\@ifundefined{textcolor}{}
{%
 \definecolor{BLACK}{gray}{0}
 \definecolor{WHITE}{gray}{1}
 \definecolor{RED}{rgb}{1,0,0}
 \definecolor{GREEN}{rgb}{0,1,0}
 \definecolor{BLUE}{rgb}{0,0,1}
 \definecolor{CYAN}{cmyk}{1,0,0,0}
 \definecolor{MAGENTA}{cmyk}{0,1,0,0}
 \definecolor{YELLOW}{cmyk}{0,0,1,0}
}

\usepackage{float}

\makeatother

\usepackage{babel}
\begin{document}

\title{Ising-like models on arbitrary graphs : The Hadamard way}

\author{Rémy Mosseri}

\address{Laboratoire de Physique Théorique de la Matière Condensée, UMR 7600
CNRS/UPMC/Sorbonne Universités, 4 place Jussieu, F-75005 Paris, France}

\email{remy.mosseri@upmc.fr}

\begin{abstract}
We propose a generic framework to describe classical Ising-like models defined on arbitrary graphs. The energy spectrum is shown to be the Hadamard transform of a suitably defined sparse "coding" vector associated with the graph. We expect that the existence of a fast Hadamard transform algorithm (used for instance in image compression processes), together with the sparseness of the coding vector may provide ways to fasten the spectrum computation. Applying this formalism to regular graphs, such as hyper cubic graphs, we obtain a simple recurrence relation for the spectrum, which significantly speeds up its determination. First attempts to analyze partition functions and transfer matrices are also presented.
\end{abstract}

\maketitle

\section{Introduction}

Ising-like interactions enter paradigmatic sets of models defined to describe phase transitions in statistical physics, and has been
applied to a very large set of a priori different problems, even outside physics. It is clearly not achievable to summarize in few
words the many contributions on that subject, and the interested reader is referred to the numerous review papers or books, see for instance Ref.\cite{toda92}.
Nevertheless useful to recall is the initial Ising-Lenz solution for the one-dimensional case\cite{ising25}, and the absence in that
case of phase transition at positive temperature, the Peierls proof for the existence of a phase transitions in two dimensions\cite{peierls36},
followed later by the Onsager solution\cite{onsager44}. Up to now, and even though a large amount of work gives precise informations
about the phase diagram in three dimensions, no exact solution is known in that case. 

In this paper, we present an approach to address  such models where classical two-level spins located at graph vertices interact
through n-body terms (regular or disordered), which we generically call "Ising-like" models, the well known standard  Ising model corresponding to one- and two-body interactions. We show how some simple algebraic tools can be used to describe and speed up the numerical solution of the
spectral problem defined on an arbitrary graph. For sake of simplicity,  we shall mainly concentrate our presentation
on the standard Ising model, keeping in mind that most of what is said applies as well in the Ising-like cases. 
In a first step, we define a binary vector which fully describe the graph structure, and a ``coding vector'' which embodies the Hamiltonian. We then show that, under an Hadamard transformation, this coding vector is sent to a ``spectral vector'', whose components are the set full set of energies in the configuration space. An important point here is to recall the existence of a ``Fast Hadamard Transform'' (FHT), able in principle to considerably speed
up the computation. 

We then concentrate on some regular graphs (mainly hyper cubes), for
which even stronger results can be derived. Finally, some preliminary
attempts to adapt this formalism to transfer matrix and partition
function determination will also be presented.

\section{Ising-like models on arbitrary graphs}

\subsection{Ising-like Hamiltonian}

We consider a graph $G$, with its $N$ vertices gathered in a set
$V$ and its bonds, as pairs of neighboring vertices, in a set $B$.
The simplest Ising model describes classical spins $S_{i}=\pm1$ located
at the $G$ vertices, subject to a constant magnetic field $h$, and
to constant pairwise interactions for spins connected by a graph edge.
The spin configuration space is the product $(\pm1)^{\otimes N}$,
forming a hyper cube $\Gamma_{N}$ in $N$ dimensions, with its $M=2^{N}$
vertices encoding the possible spin states in $G$. The Ising model
associates to each vertex in $\Gamma_{N}$ an energy :

\begin{equation}
\mathcal{E}_{\{S_{i}\}}=J\sum_{<i,j>\in B}S_{i}S_{j}\;+h\:\sum_{i\in V}S_{i},
\end{equation}

with $J<0$ (resp. $>0)$ favoring ferromagnetic (resp. antiferromagnetic)
ordering. For sake of presentation, we shall mainly discuss here the
Ising model regular case, with equal $h$ and $J$ terms for all sites
and bonds. However, most of what will be said below in the first part,
with respect to the Hadamard Transform action, applies as well to disordered
(spin glass) problems, and even to $n$-body interactions.

\subsection{Coding graphs and interacting Hamiltonians into binary vectors}

In general, a graph is rather faithfully described by its adjacency
$N\times N$ matrix $A$, with $A_{ij}=1$ whenever sites $(i,j)\in B$
and zero instead. But, especially for non-planar graphs, we may be
interested in having a finer description, pointing onto higher dimensional
cells of the graph (faces, polyhedra, polytopes, etc.). A simple generic
binary vector $\left|G\right\rangle $ of size $2^{N}$, in a space
with basis $\mathcal{B=}\left\{ 0,1\right\} ^{\otimes N}$ fulfills
this task in the following way : a basis vector is an ordered sequence
of $0$ and $1$. Each binary digit position will correspond to a
vertex of the graph; the occurrence of these different vertices in
the graph is marked by a component $1$ for the basis vectors having
one $1$ at the site position and $N-1$ zeros elsewhere. A $n$-cell
of the graph is a subset of $n$ vertices; its presence in the graph
is encoded in $\left|G\right\rangle $ as a component $1$ for the
basis vector having $n$ terms equal to $1$ at these vertices position
and zero elsewhere.

The next step is to code the interacting $n$-body spin Hamiltonian as
a coding vector $\left|C\right\rangle $. The interacting terms are
associated to $n$-body cells : the external magnetic field acts onto
one-body cells (vertices), the standard Ising interaction onto (two-body)
edges, etc ... The vector $\left|C\right\rangle $ has non vanishing
components, equal to the related $n$-body interaction term, precisely
for those basis vector which code the corresponding $n$-body cell in
$\left|G\right\rangle $. As a very simple example, a graph composed
of two vertices joined by a bond will be characterized, in the basis
$\left\{ \mid00>,\mid01>,\mid10>,\mid11>\right\} $, by $\mid G>=\left(0,1,1,1\right)$,
and its (Ising model) coding vector as $\mid C>=\left(0,\, h_{1},\, h_{2},\, J_{1,2}\right)$;
the site position is ordered here from right to left, as in standard
binary decomposition. Notice also that we are now switching from the
above $(\pm1)^{\otimes N}$ spin configuration notation to the equivalent
$\left\{ 0,1\right\} ^{\otimes N}$one with, say, an implicit bijection
between $+1$ (or $\uparrow$) to the binary label $0$.

\section{Ising spectrum from Hadamard transform}

\subsection{Hadamard Transform}

We describe now how to simply compute these interacting spin models
spectra, eventually leading to a natural speed up in their numerical
derivation. The spectrum has $M$ terms (therefore exponential with
$N)$, discretely ranging in an interval which generically grows linearly
with $N$. It is therefore, as is well known, highly degenerate, in
a way which cannot be solely captured by the graph $G$ and global
spin symmetries. We propose here to analyze this spectrum with tools
analogous to the Fourier transformation, but adapted to the peculiar
$\Gamma_{N}$ hypercube geometry. This is precisely the task fulfilled
by the so-called Hadamard transform (HT; one sometimes finds additional
names associated to it, like Walsh and Rademacher), which is a linear
transformation on a $2^{N}$-dimensional space, given, with the standard
basis $\left\{ 0,1\right\} ^{\otimes N}$, by the $2^{N}\times2^{N}$
Hadamard matrix $H_{M}$ defined recursively (with $H_{0}=1$) as
\begin{equation}
H_{m+1}\:=\:\left(\begin{array}{cc}
H_{m} & H_{m}\\
H_{m} & -H_{m}
\end{array}\right);\;\mathrm{with}\; H_{1}=\:\left(\begin{array}{cc}
1 & 1\\
1 & -1
\end{array}\right)\;\mathrm{and}\; H_{2}=\left(\begin{array}{cccc}
1 & 1 & 1 & 1\\
1 & -1 & 1 & -1\\
1 & 1 & -1 & -1\\
1 & -1 & -1 & 1
\end{array}\right)
\end{equation}

A normalization factor $1/\sqrt{2}$ is often multiplied on the right-hand-side
for $H_{1}$, but will prove useful only in the second part of this
presentation. The above relation reads also $H_{m+1}=H_{1}\otimes H_{m}$;
one easily also gets that $H_{p+q}=H_{p}\otimes H_{q}$. Hadamard
matrices have the nice property that their rows or columns are pairwise
orthogonal. The component $H_{p,q},$ with line $p$ and row $q$
both starting from zero, can be written in a compact way, by considering
the base 2 decomposition $p_{j}$ and $q_{j}$ of $p$ and $q$ :

\begin{equation}
H_{p,q}=(-1)^{\sum p_{j}q_{j}}\label{eq:Hpq}
\end{equation}

\subsection{The Hadamard transformed coding state}

For a given spin configuration, each bond contributes through a term
$\pm J$ to the energy, according to whether the two spins points
to similar or opposite directions. In addition, each site contributes
by $\pm h$ according to the local spin orientation. The complexity
in computing the spectrum amounts to adding ``coherently'' these
contributions for each $\left\{ S_{i}\right\} $ configuration. Coherent
addition of amplitudes is precisely what the quantum framework knows
how to deal with. This is why, although not a necessity, we shall
adopt quantum-like notations to describe our proposed recipe.

Let us write the Ising Hamiltonian for quantum spins (in the $S^{z}$
basis) with Pauli matrices instead of classical spins (only $\sigma^{x}$
matrices are used, so no added quantum complications)

\begin{equation}
\mathcal{H}=J\sum_{<i,j>\in B}\sigma_{i}^{x}\sigma_{j}^{x}\;+h\:\sum_{i\in V}\sigma_{i}^{x}
\end{equation}

Let $\mid0>$ be the fully polarized (which could be called here a
vacuum state) in the initial standard $\sigma^{z}$ basis. As easily
checked, the above Ising coding state simply reads $\mid C>=\mathcal{H\mid\mathrm{0}>}$. 

We now define the spectral state $\mid E>$ as the HT of $\mid C>$: $\mid E>=H_{N}$ $\mid C>$. The main first result of this paper states that $\mid E>$ have components which are precisely the energies of the associated classical Ising model we started with, a point which can be understood by considering carefully how the above
$H_{p,q}$ terms enters the matrix action in the Hadamard transformation.
Notice that, interestingly, in the quantum spin context, this amounts to a basis change from the $\sigma^{z}$ to the $\sigma^{x}$ basis (a basis change precisely done by $H_{N}$, if the above mentioned normalization factor $1/\sqrt{2}$ is introduced). $\mid C>$ is the superposition of an entangled bond vector$\mid B>=J\,\sum_{<i,j>\in B}\sigma_{i}^{x}\sigma_{j}^{x}\mid0>$ and site vector $\mid S>=h\,\sum_{i\in B}\sigma_{i}^{x}\mid0>$. The HT of a given component, say $\sigma_{i}^{x}\sigma_{j}^{x}\mid0>$, peaks out a factor $\pm1$ according to whether spins$\left(i,j\right)$
are parallel or not (see expression (3)). The contributions are then added coherently, and the total amplitude is precisely the corresponding Ising classical energy. 

Notice that the HT has already been mentioned in the context of spin glass Ising models \cite{lidar}, but used quite differently. Here, the introduction of the spectral vector $\mid E>$ as the HT of the coding state make its role direct and transparent.

To illustrate our approach, let us describe the simplest cases of two and three spins forming a bond and a triangle. With $N=2$, the coding state has already been given above, and the spectral state reads
\begin{equation}
\mid E>_{2}=\left(2h+J,-J,-J,-2h+J\right)
\end{equation}
 whose components are indeed the classical Ising energies. With $N=3$,
the coding state reads $\mid C>=\left(0,\, h,\, h,\, J,\, h,\, J,\, J,\,0\right)$
, which is transformed by $H_{3}$ to the Ising amplitudes 
\begin{equation}
\mid E>_{3}=\left(3h+J,\, h-J,\, h-J,\,-h-J,\, h-J,\,-h-J,\,-h-J,\,-3h+J\right)
\end{equation}

\subsection{Discussion}

We end this first part by a set of remarks :

\subsubsection{Fast Hadamard Transform.}

Even though interesting in itself, the present framework may not prove to be operational for spectrum determination, as compared to more straightforward enumerations, owing to the manipulation of large size Hadamard matrices. However a quite interesting feature arises from the existence of a FHT algorithm, widely used for instance in early digital image compression\cite{pratt69}, which therefore should provide ways to fasten the Ising spectrum computation for an arbitrary graph once the coding state is known. FHT relies upon the above iterative construction of $H_{m+1}$ from $H_{1}$ and $H_{m}$, which iteratively ``block diagonalizes'' the transformation. Whether the use of FHT, combined with the sparseness of the coding vector (although not evident for the above two examples with two and three spins), would significantly speed up spectral computations, and at least compare to classical methods, is certainly worth being further checked. A simple way to deal with the sparseness of $\mid C>$ is just to focus on those Hadamard matrix columns corresponding to the non vanishing elements of the coding states; such a column is readily computed as a tensor product of $(1,1)$ and $(1,-1)$ column vectors, corresponding to the ordered occurrences of ``0'' or ``1'' in the binary decomposition of the column position ( where, as for the derivation of expression \ref{eq:Hpq}, the first column start at position $0$). In the case of regular graphs, as defined below, we will find a significant speed increase through the derivation of a recursive relation for the spectrum.

\subsubsection{Extension to more complex models.}

Up to now, we have concentrated on the regular Ising model model, with constant $J$ and $h$ parameters. However, inspection of expression \ref{eq:Hpq} shows that the coherent addition of interaction contributions pertains for non regular parameters. A spin glass-like coding state, where disorder affects either the local magnetic field or the couplings (or both) will also lead to the correct spectral state. Even more, as again can be checked from expression \ref{eq:Hpq}, generalized classical spin $1/2$ models containing n-body interactions can be treated equally.

\subsubsection{Experimental implementation.}

As noted above, the HT operates a basis change from the the $\sigma^{z}$ to the $\sigma^{x}$ basis. This opens the possibility for an interesting experimental implementation : first prepare the (generically) entangled coding state in the $\sigma^{z}$ basis, and then measures this state in the $x$ orientation. Recall however that the spectral state $\mid E>$ has the Ising energies as ``probability'' amplitudes, which should be squared to compare with experimental occurrences (and therefore would not distinguish two energies opposite in sign). We do not claim here that this would give an operational method to compute large Ising models spectra. But, with an entangled coding state to be prepared, and quantum coherence to be achieved through the HT, it could already provide an interesting test for any quantum machine claiming to work in a quantum regime.

\subsubsection{Transformation paths from the coding to the spectral state.}

The standard Fourier transform relates a real and a dual reciprocal (or momentum) space. We see here that the HT (which can be viewed as a multidimensional discrete Fourier transform) relates a vector coding interactions on a graph to a (dual) vector coding the corresponding Ising energies. Suitably normalized, the HT is a unitary transformation; it is then rather tempting to follow continuously this transformation in spin space, ending at the full HT, and see what could be learned from this. Some preliminary attempts on small systems display interesting but complex patterns for the amplitudes and the degeneracies, but did not lead us to a clear or meaningful understanding of the coding state trajectory.

A different approach would consist in applying the H.T. spin after spin, following a discrete path to reach the full HT. In fact, a close inspection shows that it corresponds to the way the FHT algorithm works.

\section{Ising model on regular structures}

By a regular structure we mean here a structure which can be built
iteratively by adding identical substructures. A simple case is provided
by an hypercube $\gamma_{d}$ in dimension $d$, made of two $\gamma_{d-1}$
(note that, hypercubes being ubiquitous here, we use $"\gamma"$ for
the structure, and $"\Gamma"$ for the configuration space). We show
now that, for such regular structures, the spectral state $\mid E>$
can be computed iteratively in a rather efficient way, even compared
to a direct use of a standard FHT. The demonstration will use some algebraic
tools that are first recalled. Notice that the proposed framework
also apply, as a one-step procedure, for line to line or plane to
plane addition in a lattice and allows therefore for standard transfer
matrix constructions, which are discussed in a second step.

\subsection{Some definitions and notations}

Let us first recall some linear algebra operations, and define related
notations, that will prove useful below. All are elementary, but at
the same time refer to sometimes unusual tools (at least for a physicist)
like the  matrix Hadamard product or $\mathsf{vec}$-like operators.

We write $\mathcal{\mathscr{\mathbb{I_{\mathrm{n}}}}}$ the identity
matrix of size $n$, $\mathbb{\mathsf{1_{n}}}$ the $n\times n$ matrix
made of $1$, and $\mathbb{\mathsf{\mid1_{n}>}}$ the corresponding
size $n$ vector. For a vector $\mid X>$, we define the diagonal
matrix $D_{\left|X\right\rangle }$ whose component $D_{jj}$ is the
$j^{th}$ component $X_{j}$.

\subsubsection{Hadamard products}

The matrix Hadamard product ``$\circ$'', also called entry wise
product, multiplies pairwise elements of two matrices to form a third
one ($C=A\circ B$, such that $C_{ij}=A_{ij}B_{ij}$). We shall also
use below a ``vector Hadamard product'' $\mid A>\circ\mid B>$ whose
components are the pair-wise product of components, and a left (resp.
right) Hadamard product of a vector and a matrix, which amounts to
multiply the matrix $i^{th}$ row (resp. column) by the $i^{th}$
component of the vector. Notice that, for a matrix $M$ and vector
$\mid A>$ of same linear size, 
\begin{equation}
\mid A>\circ M=D_{\left|A\right\rangle }.M,\;\mathrm{and}\; M\circ\mid A>=M.D_{\left|A\right\rangle }\label{relation8}
\end{equation}

Having in mind the usual function elementary expansion, we shall call
$\overset{\circ}{\exp}$ the `` Hadamard matrix exponential'' whose
effect is $\left(\overset{\circ}{\exp}A\right)_{ij}=\exp\left(A_{ij}\right)$,
and extend this definition to vectors. We shall eventually use the
(trivial) relation :
\begin{equation}
\overset{\circ}{\exp}\left(\mid A>+\mid B>\right)=\overset{\circ}{\exp}(\mid A>)\circ\overset{\circ}{\exp}(\mid B>)\label{relation9}
\end{equation}

\subsubsection{A ``special'' sum for matrices and vectors}

We define special matrix and vector sums $\uplus$ as\cite{special sum}
:

\begin{equation}
\begin{cases}
\begin{array}{c}
A\uplus B=A\otimes\mathbb{\mathsf{1_{n}}}+\mathsf{1_{n}}\otimes B\\
\mid A>\uplus\mid B>=\mid A>\otimes\mid\mathsf{1}>+\mid\mathsf{1}>\otimes\mid B>,
\end{array}\end{cases}\label{matrix sum}
\end{equation}
where the subscript $n$ (the vector size) has been omitted, and $"\otimes"$
is the standard Kronecker product . Be careful not to confuse with
the standard matrix Kronecker sum $A\oplus B=A\otimes\mathsf{\mathfrak{\mathbb{I}}}_{\mathrm{n}}+\mathsf{\mathfrak{\mathbb{I}}}_{\mathrm{n}}\otimes B$
where the identity matrix $\mathsf{\mathfrak{\mathbb{I}}}_{\mathrm{n}}$
appears instead the ``one matrix'' $\mathsf{1}.$ Notice the useful
relation :

\begin{equation}
\overset{\circ}{\exp}\left(A\uplus B\right)=\overset{\circ}{\exp A}\otimes\overset{\circ}{\exp B},\label{relation7-1}
\end{equation}

which reminds (but differs from) the better known $\exp(A\oplus B)=\exp A\otimes\exp B$
involving the standard matrix exponential and Kronecker matrix sum.

\subsubsection{The $\mathsf{vec}$ operator}

The $\mathsf{vec}$ operator maps a $n\times m$ matrix onto a column
vector of size $n*m$, simply built by stacking the successive columns
of the original matrix. We are mainly interested here by square matrices,
and would like also to use the inverse of a $\mathsf{vec}$ map, unraveling
a vector of size $n^{2}$ onto an $n\times n$ matrix with its successive
columns taken from the vector components. There is an apparent notation
conflict here, some authors calling $\mathsf{devec}$ this operator,
while others\cite{turkington} use $\mathsf{devec}$ to denote the
operator building a row vector of size $n*m$ from the original matrix
by concatenation of its successive rows . Therefore we shall adopt
here the notation $\mathsf{unvec}$ for the inverse of $\mathsf{vec}.$
The following useful identity, involving three matrices $P$, $Q$
and $R$ will be used later:
\begin{equation}
\mathsf{vec}(P.Q.R)=(R^{T}\otimes P).\mathsf{vec(Q),}\label{vec}
\end{equation}

where $R^{T}$ is the matrix transpose of $R$, and the ``dot''
refers to the standard product for matrix algebra (between two matrices
or a matrix and a vector).

\subsection{The hypercube case}

We now discuss a recursive algorithm to compute the Ising model spectra
for the family of hypercubes with increasing dimensionality. Notice
that, interestingly, a hypercube $\gamma_{2d}$ is topologically equivalent
to an hypercubic lattice of lateral size four in dimension $d$, with
periodic boundary conditions (PBCs), noted here $Z_{d}^{4}$. The latter
inherits the former symmetry group, of order $(2d)!2^{2d}$. The hypercube
$\gamma_{d}$ has $N=2^{d}$ sites; its spin configuration space is
the hyperbole $\Gamma_{M}$ with $M=2^{N}$. The recursive additive
construction $\gamma_{d+1}=\gamma_{d}\cup T(\gamma_{d})$ , where
$T(\gamma_{d})$ is a translated version of $\gamma_{d}$ in the new
space dimension, leads to a Kronecker product for the respective configuration
space : $\Gamma_{M^{2}}=\Gamma_{M}\otimes\Gamma_{M}$.

\subsubsection{Ising addition spectrum for two uncorrelated subgraphs }

Let us start by trivially writing the spectral vector for the union
of two graphs $G_{1}$ and $G_{2}$, both with $N$ sites for sake
of simplicity, but with possibly different interaction coding states
(Ising model parameters). Compute first (say through FHT) the two
spectral states $\mid E_{1}>$ and $\mid E_{2}>$. If the two graphs
have no interactions, the spectral state of their union $G$, noted
$\mid E>$ , is simply the ``addition spectrum'' of its two parts,
an object which is precisely given by the above defined special vector
sum $\uplus$ : $\mid E>\:\begin{aligned}\end{aligned}
=\,\mid E_{1}>\uplus\mid E_{2}>$. 

The latter expression can be easily checked directly; let us nevertheless
derive it in the present framework. Call $\left|0\right\rangle _{N}$
the fully polarized state with $N$ spins, and $\left|C_{1}\right\rangle $
and $\left|C_{2}\right\rangle $ the coding states associated to $G_{1}$
and $G_{2}$. Since up to now the latter are not Ising connected,
the coding state for $G$ simply reads
\begin{equation}
\left|C\right\rangle =\left|C_{2}\right\rangle \otimes\left|0\right\rangle _{N}+\left|0\right\rangle _{N}\otimes\left|C_{1}\right\rangle .\label{eq:productgraph}
\end{equation}
$\left|0\right\rangle _{N}$ is the first element in the standard
basis $\left\{ 0,1\right\} ^{\otimes N}$, and its HT
is easily shown to be $\left|1\right\rangle _{N}$ . As a result,
the spectral state $\left|E\right\rangle $ reads
\begin{equation}
\left|E\right\rangle =H_{2N}.\left|C\right\rangle =(H_{N}\otimes H_{N}).\left|C\right\rangle =\mid E_{2}>\otimes\left|1\right\rangle _{N}+\left|1\right\rangle _{N}\otimes\mid E_{1}>=\mid E_{2}>\uplus\mid E_{1}>
\end{equation}

Now really interesting questions start when the graphs union is dressed
with new couplings between the parts, with the task of combining coherently
these new interactions with the simple ``disconnected'' addition
spectrum.

\subsubsection{Introducing interactions between the subgraphs}

There are clearly three types of Ising (bond) interactions : those
associated separately to $G_{1}$ or $G_{2}$, and those connecting
the two graphs. For reasons which will be clear below, we choose to
consider the matrix $C_{2N}=\mathsf{unvec}\,(\mid C_{2N}>)$. The
first two types of bonds appear in its first column as the column
vector $\mid C_{1}>$, and in its first row as the line vector $<C_{2}\mid$,
a direct consequence of the above uncorrelated graph coding state
in formula (\ref{eq:productgraph}). The remaining part, forming a
matrix $D$, codes the remaining interactions. It has non vanishing
elements only at positions corresponding to an interacting bond between
the graphs $G_{1}$ and $G_{2}$. At this stage, this description
is still generic and applies to the union of two different graphs,
with possibly non regular bonding between the two graphs, the complexity
being coded in $D$. We now show that for the class of regular graphs
to which hypercubes belong, $D$ takes a simple form which allow for
a recursive analysis.

\subsubsection{Ising spectrum for the hypercube}

The additive construction of hypercubes translates into a simple iterative
construction for the coding state vector from  $\mid C_{N}>$ to $\mid C_{2N}>$.
If we are cautious enough to number the sites coherently for the two
$\gamma_{d}$ copies (which means sites $j$ and $j+N$ are translation
related), we get a rather simple form for the matrix $D$ which turns
into a pure diagonal matrix, the third type of bonds contributions
sitting at location $2^{j}+1$ along the diagonal, with integer $j$
running from $0$ to $N-1$, and being simply related to the above
defined site vector $\mid S_{N}>$ associated to $\gamma_{d}$ : $D_{2N}=J\, D_{\left|S_{N}\right\rangle }$.

To clarify our notations, let us display these objects for the easiest
case, from the point $\gamma_{0}$ to the segment $\gamma_{1}$ and
then to the square $\gamma_{2}$ : 
\begin{equation}
\mid C_{1}>=\left(\begin{array}{c}
0\\
h
\end{array}\right),\;\mid C_{2}>=\left(\begin{array}{c}
0\\
h\\
h\\
J
\end{array}\right),\; C_{2}=\left(\begin{array}{cc}
0 & h\\
h & J
\end{array}\right),\; D_{2}=\left(\begin{array}{cc}
0 & 0\\
0 & J
\end{array}\right),\; C_{4}=\left(\begin{array}{cccc}
0 & h & h & J\\
h & J & 0 & 0\\
h & 0 & J & 0\\
J & 0 & 0 & 0
\end{array}\right),\; D_{4}=\left(\begin{array}{cccc}
0 & 0 & 0 & 0\\
0 & J & 0 & 0\\
0 & 0 & J & 0\\
0 & 0 & 0 & 0
\end{array}\right)\label{B4}
\end{equation}

We aim to compute the Ising spectrum vector $\mid E_{2N}>=H_{2N}\mid C_{2N}>$.
This could be done easily using FHT as discussed above. But we proceed
differently here, by recalling that $H_{2N}=H_{N}\otimes H_{N}$ and
that $H_{N}^{T}=H_{N}$. We than use relation (\ref{vec}), with $P=R=H_{N}$
and the above defined $C_{2N}$ to get :
\begin{equation}
\mid E_{2N}>=\mathsf{vec}\left(H_{N}.C_{2N}.H_{N}\right)\label{E_2N}
\end{equation}

This only difficult part requires now to evaluate a set of matrices
$M_{2N}=H_{N}.D_{2N}.H_{N}$, the Hadamard transformed $D_{2N}$ matrices,
which happens to follow a simple iterative construction. Some easy
algebra leads to the following coupled system, which iteratively constructs
the hypercube Ising spectrum in any dimension :

\begin{align}
\begin{cases}
\begin{array}{c}
\mid E_{2N}>\:\begin{aligned}\end{aligned}
=\,(\mid E_{N}>\uplus\mid E_{N}>)+\,\mathsf{vec}\left(M_{2N}\right)\\
M_{2N}\,\begin{aligned}\end{aligned}
=\, M_{N}\uplus M_{N}
\end{array}\end{cases}\label{E_2Nbis}
\end{align}

As already said, the first term on the RHS of the first equation provides
the expected addition spectrum from the two copies of $\gamma_{n}$,
while the second term contains their interaction. We leave as an exercise
to the interested reader to recover, from the initial $\mid E_{1}>=\left(h,-h\right)$
and $M_{2}=\mathsf{unvec}\left(J,-J,-J,J\right)$, the Ising spectrum
for a square :

\begin{equation}
\mid E_{4}>=(4h+4J,2h,2h,0,2h,0,-4J,-2h,2h,-4J,0,-2h,0,-2h,-2h,-4h+4J)
\end{equation}

Relation (\ref{E_2Nbis}) is very easily iterated numerically. As
an example, two additional iterations leads, almost instantaneously
on a standard computer, to $\mid E_{16}>$ , the spectrum for $\gamma_{4}$,
and therefore for the $4\times4$ square lattice system with periodic
boundary conditions. It reads, with $J=1$ and $h=0$, given in frequency
(pairs of numbers, first term for the energy, and second term its
degeneracy) :

\begin{equation}
\left\{ \left(-32,2\right),\left(-24,32\right),(-20,64),\left(-16,424\right),\left(-12,1728\right),\left(-8,6688\right),\left(-4,13568\right),\left(0,20524\right),\ldots\right\} \label{spectre gamma4}
\end{equation}
We omit the positive part of the spectrum, the latter being symmetrical.
Notice however that iterating relation (\ref{E_2Nbis}) involves consecutive
tensor products, which is paid for in terms of computer memory. If the
$\gamma_{5}$ spectrum (which corresponds to a piece of cubic lattice
of size $4\times4\times2$) is still very quickly computed, higher
sizes turn out to be difficult to reach along this direct track, if
 equipped only with standard computers memory.

\subsubsection{Partition function}

We now aim to evaluate the partition function
\begin{equation}
\mathcal{Z}=\sum_{\left\{ S_{i}\right\} }\exp\left(-\beta\mathcal{E}_{\left\{ S_{i}\right\} }\right),\label{partition}
\end{equation}
from which all interesting thermodynamical properties can be derived.
The reader who has followed the present approach might have already
guessed that we shall define a partition function vector $\mid Z>$
as :
\begin{equation}
\mid Z>=\overset{\circ}{\exp}\left(-\beta\mid E>\right)\label{vectorZ}
\end{equation}

where we use the above defined matrix Hadamard exponential. The standard
$\mathcal{Z}$ is simply the sum of $\mid Z>$ components. It is again
not difficult to derive an iterative relation to get $\mid Z_{2N}>$
from $\mid Z_{N}>$, equivalent to expression (\ref{E_2Nbis}) we
had for the spectrum. With the diagonal matrix $D_{\left|Z\right\rangle }$
built from $Z_{N}$ and $W_{N}=\overset{0}{\exp}(-\beta M_{N})$,
it reads 
\begin{flalign}
\begin{cases}
\begin{array}{c}
\mid Z_{2N}>\:\begin{aligned}\end{aligned}
=D_{\left|Z\right\rangle }^{\otimes2}.\mathsf{vec}\left(W_{2N}\right)=\,\mathsf{vec(}D_{\left|Z\right\rangle }.W_{2N}.D_{\left|Z\right\rangle })\\
W_{2N}\,\begin{aligned}\end{aligned}
=\, W_{N}\otimes W_{N}
\end{array}\end{cases}\label{Z_2N}
\end{flalign}

Although quite simple, we detail, in Appendix A, the step by step
passage from relation $\left(\ref{E_2Nbis}\right)$ to relation $\left(\ref{Z_2N}\right)$.
The second line of relation \ref{Z_2N} arises directly from that
of relation \ref{E_2Nbis} upon using relation (\ref{relation7-1}).
Quite interestingly, upon additional manipulations, the partition
function itself, $\mathcal{Z}_{2N}$, which is the sum of all $\mid Z_{2N}>$
elements, simply reads 
\begin{equation}
\mathcal{Z}_{2N}=<Z_{N}\mid W_{2N}\mid Z_{N}>\label{partition_2N}
\end{equation}

Let us stress that the present approach is different from that followed
in Ref.\cite{vendennest07}, where the partition function $\mathcal{Z}$
was computed as the inner product of a stabilizer state and a product
state, and for the sake of which these authors defined spins living
on edges of the graph. In the present case, the square partition function
$\mathcal{Z}_{4}$ is computed as, with $x=\exp\left(-\beta J\right)$,
\begin{equation}
\mathcal{Z}_{4}=\begin{array}{cccc}
(x, & x^{-1}, & x^{-1}, & x)\end{array}\left(\begin{array}{cccc}
x^{2} & 1 & 1 & x^{-2}\\
1 & x^{2} & x^{-2} & 1\\
1 & x^{-2} & x^{2} & 1\\
x^{-2} & 1 & 1 & x^{2}
\end{array}\right)\left(\begin{array}{c}
x\\
x^{-1}\\
x^{_{_{-1}}}\\
x
\end{array}\right)=2(6+x^{4}+x^{-4})\label{Z4_1}
\end{equation}
 An interesting point arises here. The matrix $W_{N}$ , which encodes
the coupling between the two sub parts, is diagonalized by a Hadamard
matrix (now normalized to have unitary transformations). Indeed, start
with the first in the series, 
\begin{equation}
W_{2}=\left(\begin{array}{cc}
x & x^{-1}\\
x^{-1} & x
\end{array}\right)
\end{equation}

It is diagonalized under transformation by $H_{1}$:
\begin{equation}
W_{2}^{'}=H_{1}.\, W_{2}\,.\, H_{1}=\left(\begin{array}{cc}
x+x^{-1} & 0\\
0 & x-x^{-1}
\end{array}\right)\quad\mathrm{with}\:\mathrm{now}\; H_{1}=\frac{1}{\sqrt{2}}\left(\begin{array}{cc}
1 & 1\\
1 & -1
\end{array}\right)
\end{equation}

Due to the simple tensor product construction of $W_{N}$ , explicit
in the second line of $\left(\ref{Z_2N}\right)$, together with the
tensor product construction for the Hadamard matrices, the spectrum
of $W_{N}$ is computed directly from that of $W_{2}$. Recalling
that Hadamard matrices square to identity, we can insert the identity
matrix on the left and on the right of $W_{2N}$ in expression $\left(\ref{partition_2N}\right)$.
This leads to an expression for the partition function as the expectation
value of the (known) diagonalized form of $W_{2N}$ , taken on the
HT of the partition function of the sub parts $\mid Z_{N}>$.
For the square case , expression $\left(\ref{Z4_1}\right)$ now rewrites
: 

\begin{equation}
\mathcal{Z}_{4}=\begin{array}{cccc}
(x+x^{-1}, & 0, & 0, & x-x^{-1})\end{array}\left(\begin{array}{cccc}
x^{2}+x^{-2}+2 & 0 & 0 & 0\\
0 & x^{2}-x^{-2} & 0 & 0\\
0 & 0 & x^{2}-x^{-2} & 0\\
0 & 0 & 0 & x^{2}+x^{-2}-2
\end{array}\right)\left(\begin{array}{c}
x+x^{-1}\\
0\\
0\\
x-x^{-1}
\end{array}\right)=2(6+x^{4}+x^{-4})\label{Z4_2}
\end{equation}

Notice that for the case of two-dimensional grids, and without disorder, there are powerful methods to derive partition functions for large but finite size systems\cite{klas}. 

\subsection{Transfer matrices}

As already mentioned, the above step by step treatment for the hypercubes
also applies, as a one-step procedure, to describe line to line or
plane to plane (or graph to graph) addition to form a repeated structure
in a given direction. Call $\Lambda_{N}$ the graph with $N$ vertices
to be repeated, and $\Lambda_{2N}=\Lambda_{N}\cup T(\Lambda_{N})$
the resulting graph after one step. We consider the simplest case
where each vertex in $\Lambda_{N}$ has one (translated) neighbor
in $T(\Lambda_{N})$, suitably numbered as before. The case where
each vertex has several new neighbors in the translated copy (as
occurs for instance in dense lattices like fcc) could also be treated,
with some complications, and will not analyzed here. With $\mid E_{N}>$
and $\mid Z_{N}>$ for the Ising spectrum and partition vector for
the $\Lambda$ part, relations (\ref{E_2Nbis}) and (\ref{Z_2N})
still apply to get $\mid E_{2N}>$ and $\mid Z_{2N}>$. But, more
interestingly, and as expected, the transfer matrix itself is easily
derived. $T_{N}$ , the $2^{N}\times2^{N}$ transfer matrix from $\Lambda_{N}$
to $T(\Lambda_{N})$, simply reads :
\begin{equation}
T_{N}=W_{2N}.\, D_{\left|Z\right\rangle }=W_{2N}\circ\mid Z_{N}>\label{transfer matrix}
\end{equation}

With $m$ copies of $\Lambda$ (and periodic boundary condition) ,
the partition function follows from the well known trace relation
$\mathcal{Z}_{mN}=Tr\left(T_{N}^{m}\right)$.

While a full diagonalization of $T$ would solve the problem, we shall
ask, more modestly, whether Hadamard transformations (with normalized
Hadamard matrices from now on) can be helpful to further simplify
$T_{N}$, under the form of a partial block-diagonalization. A simple
first step, but nevertheless rather trivial, consists in operating
a global HT to the transfer matrix; notice that one
should better say an Hadamard conjugation since it reads : $T\rightarrow H^{-1}.T.H$.
This leads to split the transfer matrix into two separate blocks,
which in fact translates the existence of a global spin-flip symmetry.
One could then try to go further and operate with lower order Hadamard
matrices on the two blocks. First attempts on small systems show that
it might be the case, although no precise conclusions could yet be
drawn. As an example, we detail the process in Appendix B for the
$16\times16$ transfer matrix $T_{4}$ connecting two rings of four
spins, which we analyze here in the vanishing $h$ magnetic field
limit, and for which a partial block diagonalization has been achieved.

\section{Conclusion}

We have presented a general framework to address Ising-like models on arbitrary graphs. The interactions between spins are neither limited to pairwise interactions, nor restricted to constant couplings. The present approach relies on a dual structure : a coding state on one side, which represents the interactions carried by the spins, is transformed into a spectral state whose amplitudes are the model energies. The transformation is operated by Hadamard matrices. 

A interesting point here is the existence of a FHT, which provides therefore potential for speed increase for numerical computation, in particular if sparse vectors manipulations are cleverly introduced. Whether FHT, sparse algebra and parallelism could be efficiently associated here is an open question that may interest numerical experts.

When the graph is obtained as the union of subgraphs, the proposed analysis simply separates the contribution of the sub graph spectra from that arising from their interaction. Whenever the interaction takes a regular form, we have shown how to compute iteratively the spectrum.

Attempts to compute the partition function and transfer matrices have also been described. Further works need to be done, in particular to check possible transfer matrix block diagonalization algorithms.

In addition, the fact that the HT is a main tool for quantum computation raises the question of a possible experimental implementation of the proposed approach.

Finally, one may ask whether this type of approach could be used for other collective models, like higher spins systems or the Potts model. However, using standard Hadamard matrices (with +1 and -1 entries), combined with the above defined coding vector, restricts the application to spins carrying only two distinct values (with symmetric interactions), therefore to Ising spins only.

\medskip{}

The author would like to thank J.M. Maillard for stimulating discussions, and acknowledge comments from T. Barthel and  T. Schlittler.

\section{Appendix}

\subsection{Derivation of the partition function vector $\mid Z>$ in expression
\ref{Z_2N}}

As said above, it is not difficult to derive an iterative relation
to get $\mid Z_{2N}>$ from $\mid Z_{N}>$, given in expression \ref{Z_2N}.
It provides nevertheless an interesting exercise to manipulate some
of the algebraic tools introduced in the text, and we therefore detail
it now, step by step: 

\begin{flalign*}
 & \mid Z_{2N}>\overset{def}{=}\:\overset{\circ}{\exp}\left(-\beta\mid E_{2N}>\right)\\
 & =\overset{\circ}{\exp}\left[-\beta\,(\mid E_{N}>\uplus\mid E_{N}>)-\beta\,\mathsf{vec}\left(M_{2N}\right)\right]\\
 & =\overset{\circ}{\exp[-\beta}(\mid E_{N}>\uplus\mid E_{N}>)]\circ\overset{\circ}{\exp}\left[\mathsf{-\beta\, vec}\left(M_{2N}\right)\right]\;\;\mathrm{from\:}\mathrm{(\ref{relation9})}\\
 & =[\overset{\circ}{\exp}(-\beta\mid E_{N}>)\otimes\overset{\circ}{\exp}(-\beta\mid E_{N}>)]\circ\mathsf{vec}\left(\overset{}{\overset{\circ}{\exp}\left(-\beta M_{2N}\right)}\right)\;\mathbf{\;\mathrm{from}\:(\ref{relation7-1})}\:\mathrm{and}\:\mathrm{obvious\:\mathsf{vec}/\overset{\circ}{\exp}\: commutation}\\
 & =\mid Z_{N}>^{\otimes2}\circ\:\mathsf{vec}\left(W_{2N}\right)\\
 & =D_{Z}^{\otimes2}.\:\mathsf{vec}\left(W_{2N}\right)\;\mathrm{\mathrm{\; from\:(8)}\;\;\left(Q.E.D.\right)}\\
 & =\,\mathsf{vec(}D_{Z}.W_{2N}.D_{Z})\;\mathrm{\; from}\:\mathrm{(\ref{vec})}
\end{flalign*}

The second line of relation \ref{Z_2N} arises directly from that
of relation \ref{E_2Nbis} upon using relation (\ref{relation7-1})

\subsection{Partial block diagonalization of the transfer matrix:  A simple example.}

As announced in the text, we present here the case of the $16\times16$ transfer matrix $T_{4}$ connecting two rings of four spins, which is analysed in the vanishing $h$ magnetic field limit. Notice that this simple example is just meant here to introduce a proposition which clearly need to be further checked for larger cases. 
 Using relations \ref{Z_2N} and \ref{transfer matrix}, one easilly writes $T_{4}$
as : 
\begin{equation}
T_{4}=\left(\begin{array}{cccccccccccccccc}
x^{8} & x^{6} & x^{6} & x^{4} & x^{6} & x^{4} & x^{4} & x^{2} & x^{6} & x^{4} & x^{4} & x^{2} & x^{4} & x^{2} & x^{2} & 1\\
x^{2} & x^{4} & 1 & x^{2} & 1 & x^{2} & \frac{1}{x^{2}} & 1 & 1 & x^{2} & \frac{1}{x^{2}} & 1 & \frac{1}{x^{2}} & 1 & \frac{1}{x^{4}} & \frac{1}{x^{2}}\\
x^{2} & 1 & x^{4} & x^{2} & 1 & \frac{1}{x^{2}} & x^{2} & 1 & 1 & \frac{1}{x^{2}} & x^{2} & 1 & \frac{1}{x^{2}} & \frac{1}{x^{4}} & 1 & \frac{1}{x^{2}}\\
1 & x^{2} & x^{2} & x^{4} & \frac{1}{x^{2}} & 1 & 1 & x^{2} & \frac{1}{x^{2}} & 1 & 1 & x^{2} & \frac{1}{x^{4}} & \frac{1}{x^{2}} & \frac{1}{x^{2}} & 1\\
x^{2} & 1 & 1 & \frac{1}{x^{2}} & x^{4} & x^{2} & x^{2} & 1 & 1 & \frac{1}{x^{2}} & \frac{1}{x^{2}} & \frac{1}{x^{4}} & x^{2} & 1 & 1 & \frac{1}{x^{2}}\\
1 & x^{2} & \frac{1}{x^{2}} & 1 & x^{2} & x^{4} & 1 & x^{2} & \frac{1}{x^{2}} & 1 & \frac{1}{x^{4}} & \frac{1}{x^{2}} & 1 & x^{2} & \frac{1}{x^{2}} & 1\\
\frac{1}{x^{4}} & \frac{1}{x^{6}} & \frac{1}{x^{2}} & \frac{1}{x^{4}} & \frac{1}{x^{2}} & \frac{1}{x^{4}} & 1 & \frac{1}{x^{2}} & \frac{1}{x^{6}} & \frac{1}{x^{8}} & \frac{1}{x^{4}} & \frac{1}{x^{6}} & \frac{1}{x^{4}} & \frac{1}{x^{6}} & \frac{1}{x^{2}} & \frac{1}{x^{4}}\\
\frac{1}{x^{2}} & 1 & 1 & x^{2} & 1 & x^{2} & x^{2} & x^{4} & \frac{1}{x^{4}} & \frac{1}{x^{2}} & \frac{1}{x^{2}} & 1 & \frac{1}{x^{2}} & 1 & 1 & x^{2}\\
x^{2} & 1 & 1 & \frac{1}{x^{2}} & 1 & \frac{1}{x^{2}} & \frac{1}{x^{2}} & \frac{1}{x^{4}} & x^{4} & x^{2} & x^{2} & 1 & x^{2} & 1 & 1 & \frac{1}{x^{2}}\\
\frac{1}{x^{4}} & \frac{1}{x^{2}} & \frac{1}{x^{6}} & \frac{1}{x^{4}} & \frac{1}{x^{6}} & \frac{1}{x^{4}} & \frac{1}{x^{8}} & \frac{1}{x^{6}} & \frac{1}{x^{2}} & 1 & \frac{1}{x^{4}} & \frac{1}{x^{2}} & \frac{1}{x^{4}} & \frac{1}{x^{2}} & \frac{1}{x^{6}} & \frac{1}{x^{4}}\\
1 & \frac{1}{x^{2}} & x^{2} & 1 & \frac{1}{x^{2}} & \frac{1}{x^{4}} & 1 & \frac{1}{x^{2}} & x^{2} & 1 & x^{4} & x^{2} & 1 & \frac{1}{x^{2}} & x^{2} & 1\\
\frac{1}{x^{2}} & 1 & 1 & x^{2} & \frac{1}{x^{4}} & \frac{1}{x^{2}} & \frac{1}{x^{2}} & 1 & 1 & x^{2} & x^{2} & x^{4} & \frac{1}{x^{2}} & 1 & 1 & x^{2}\\
1 & \frac{1}{x^{2}} & \frac{1}{x^{2}} & \frac{1}{x^{4}} & x^{2} & 1 & 1 & \frac{1}{x^{2}} & x^{2} & 1 & 1 & \frac{1}{x^{2}} & x^{4} & x^{2} & x^{2} & 1\\
\frac{1}{x^{2}} & 1 & \frac{1}{x^{4}} & \frac{1}{x^{2}} & 1 & x^{2} & \frac{1}{x^{2}} & 1 & 1 & x^{2} & \frac{1}{x^{2}} & 1 & x^{2} & x^{4} & 1 & x^{2}\\
\frac{1}{x^{2}} & \frac{1}{x^{4}} & 1 & \frac{1}{x^{2}} & 1 & \frac{1}{x^{2}} & x^{2} & 1 & 1 & \frac{1}{x^{2}} & x^{2} & 1 & x^{2} & 1 & x^{4} & x^{2}\\
1 & x^{2} & x^{2} & x^{4} & x^{2} & x^{4} & x^{4} & x^{6} & x^{2} & x^{4} & x^{4} & x^{6} & x^{4} & x^{6} & x^{6} & x^{8}
\end{array}\right)
\end{equation}

A first (trivial) block-diagonalization (two $8\times8$ blocks), associated
with the global spin-flip symmetry, is provided by transforming $T_{4}$
to $H_{4}.T_{4}.H_{4}$ (recall that $H_{4}=H_{4}^{-1}$ in the normalized
case). Combining suitable columns and row permutations, and Hadamard
matrices of lower order, one then gets the following spectrum for
$T_{4}$ : a set of eight eigenvalues, shown as pairs (eigenvalue, degeneracy
$d$): $\left\{ \left(x^{4}-2+x^{-4},\: d=4\right),\,\left(\left(1+x^{2}\right)\left(x^{2}-1\right)^{3}/x^{4},\: d=2\right),\left(\,\left(1+x^{2}\right)^{3}\left(x^{2}-1\right)/x^{4},\: d=2\right)\right\} $;
the following two $2\times2$ matrices :
\begin{equation}
\left(\begin{array}{cc}
x^{8}-1 & 2x^{2}\left(x^{4}-1\right)\\
\frac{2\left(x^{4}-1\right)}{x^{2}} & \frac{x^{8}-1}{x^{4}}
\end{array}\right),\;\mathrm{\; and\;\;}\left(\begin{array}{cc}
1-x^{-8} & \frac{2(x^{4}-1)}{x^{6}}\\
\frac{2\left(x^{4}-1\right)}{x^{2}} & \frac{1-x^{-8}}{x^{4}}
\end{array}\right),\;
\end{equation}

and one $4\times4$ matrix :
\begin{equation}
\left(\begin{array}{cccc}
x^{8}+1 & 2x^{4} & \sqrt{2}\left(x^{2}+\sqrt{2}x^{4}+x^{6}\right) & \sqrt{2}\left(-x^{2}+\sqrt{2}x^{4}-x^{6}\right)\\
2x^{-4} & x^{-8}+1 & \frac{\sqrt{2}\left(x^{2}+\sqrt{2}x^{4}+x^{6}\right)}{x^{8}} & \frac{\sqrt{2}\left(-x^{2}+\sqrt{2}x^{4}-x^{6}\right)}{x^{8}}\\
2+\frac{\sqrt{2}\left(1+x^{2}\right)}{x^{4}} & 2+\frac{\sqrt{2}\left(1+x^{2}\right)}{x^{4}} & x^{-4}+2\sqrt{2}x^{-2}+4+2\sqrt{2}x^{2}+x^{4} & -2\\
2-\frac{\sqrt{2}\left(1+x^{2}\right)}{x^{4}} & 2-\frac{\sqrt{2}\left(1+x^{2}\right)}{x^{4}} & -2 & x^{-4}-2\sqrt{2}x^{-2}+4-2\sqrt{2}x^{2}+x^{4}
\end{array}\right)
\end{equation}

These small matrices could again be further simplified, but (apparently)
not with HTs, so we do not show it. The interested
reader can easily recover, from $Tr\left(T_{4}^{4}\right)$, the partition
function for a $4\times4$ square lattice with PBC, whose Ising spectrum
was already given in relation (\ref{spectre gamma4}).

\begin{thebibliography}{1}
\bibitem{special sum}The present author did not find a reference,
nor a name, to quote for this rather simple type of sum, and will
appreciate suggestions
\end{thebibliography}
\end{document}